\documentclass{ws-ijmpa}
\usepackage[super,compress]{cite}
\begin{document}

\markboth{J. L.L\'opez, O. Obreg\'on and M. Sabido}
{Toward a Supergravity Spectral Action}

\catchline{}{}{}{}{}
%
%%%%%%%%%%%%%%%%%%%%%%%%%%%%%%%%%%%%%%%%%%%%%%%%%%%%%%%%%%%%%%%%%%%%

\title{Toward a Supergravity Spectral Action}

\author{J. L. L\'opez$^a$\footnote{jl\_lopez@fisica.ugto.mx} ,  O. Obreg\'on$^a$\footnote{octavio@fisica.ugto.mx} , M. P. Ryan$^b$\footnote{ryanmex2002@yahoo.com}   and M. Sabido$^a$\footnote{msabido@fisica.ugto.mx} }

\address{$^a$ Departamento de F\'{\i}sica, DCI-Campus Le\'on, Universidad de 
Guanajuato,\\
A.P. E-143, C.P. 37150,  Guanajuato, M\'exico.\\
$^b$ Instituto de Ciencias Nucleares Universidad Nacional Aut\'onoma de M\'exico,\\
 A.P. 70-543,  M\'exico D.F. 04510,M\'exico.\\
 Present Address: 30101 Clipper Lane,\\
Millington MD 21651 USA.}

\maketitle

\begin{abstract}
A spectral action for a generalized bosonic sector corresponding to the Dirac operator of Euclidean supergravity is proposed. We 
calculate, up to $a_4$, the Seeley-Dewitt coefficients in the expansion  of the spectral action. It is in general not known how to construct a ``matter fermionic'' supersymmetric partner 
to the spectral action. The action we propose provides the effective action to be completed to get, at any order of the expansion, the corresponding supersymmetric action.

\end{abstract}

\ccode{PACS numbers:04.65.+e,02.40.Gh,11.10.Nx}

\section{Introduction}	

The equivalence principle and gauge invariance are the fundamental pillars of the two most successful
theories in physics, general relativity and Yang-Mills theory. They lead to a greater understanding of the basic interactions 
in our Universe. However, these theories  seem to be incompatible at the quantum level. 
This incompatibility might suggest that they are theories arising from some other more fundamental 
principle. One of the most interesting proposals in the literature is the spectral action of 
noncommutative geometry. It involves a spectral geometry consistent with the physical measurements of distances. The usual emphasis on the points 
$x \in \mathcal{M}$ on a geometric space 
is replaced by the spectrum $\Sigma$ of the Dirac operator $D$ and it is assumed that the spectral action 
depends only on $\Sigma$. This is the spectral action principle. The spectrum is a geometric invariant 
that replaces diffeomorphism invariance. By applying this basic principle to the noncommutative 
geometry defined by the standard model, it has been shown \cite{Connes} that the dynamics of all interactions, 
including gravity, are given by the spectral action. Its heat kernel expansion in terms of the 
Seeley-DeWitt coefficients $a_n$ gives an effective action up to the coefficient considered. 
Assuming the Riemannian spin connection for the gravitational sector of the spectral 
action, the first three terms in the expansion correspond to a constant, the usual 
Einstein-Hilbert action plus Weyl gravity and a Gauss-Bonnet topological invariant. As is well known, if one would apply a Palatini formalism to this effective action
one would get a non-Riemannian connection and this result will also be different at any desired order in the expansion, namely for each effective action. \\

The Dirac operator is an essential element in the physical action of noncommutative geometry. It encodes, together with
representation of the algebra of coordinates, both geometry and physics. 
It was stated in Ref. \citen{Rovelli} that in the case of gravity one can consider the eigenvalues of the Dirac operator as observables if they satisfy certain constraints 
that restrict the phase space and the structure of the space-time manifold. The same type of analysis was later performed for Euclidean supergravity \cite{Vancea1,Vancea2} where 
also the eigenvalues of the appropriate Dirac operator can be understood as observables that must satisfy a set of generalized constraints. We were motivated by this last result and   
in this work we calculate the first three terms of the spectral action restricting our calculations to the Dirac operator of simple Euclidean supergravity. 
As is well known, the spin connection of this operator corresponds to the standard Riemannian connection, by means of which the gravitational sector of the spectral action is usually 
constructed, plus a contortion term due to the presence of the Rarita-Schwinger field.  
First, in Sec. 2 we 
review the well known \cite{Connes} spectral action construction related to pure gravity based on the standard Riemannian connection. Then in Sec. 3 we present the calculation 
of the first three Seeley-DeWitt coefficients of the 
heat kernel expansion based on the trace of the square of the supergravity Dirac operator. This procedure, however, will not provide in general a supersymmetric effective action 
at any desired order in the expansion \cite{17}. As is well known, a supersymmetric ``matter fermionic'' partner to the spectral action is in general not known. We are providing the procedure 
to obtain the bosonic part of the spectral action, here we calculate it up to the $a_4$ coefficient. We will get an ``effective generalized bosonic action''. Based on our proposal at certain order 
in the expansion one could, in principle, supersymmetrize the effective action of interest to get the appropriate ``matter fermionic'' terms. This is a complicated task and, not knowing 
a general ``matter fermionic'' action, the calculation must be performed at each desired order in the expansion, namely for each effective action independently. \\ \\
We will not attempt to complete the supersymmetry, up to the $a_4$ Seeley-DeWitt coefficient, corresponding to the ``generalized spectral bosonic action'' we propose. This is a specific particular 
task for each order in the expansion. It will be shown that the effective generalized bosonic action has the curvature term in the first order formalism and we know, at this level in the expansion, 
that we should add the Rarita-Schwinger action to get $\mathcal{N} = 1$ supergravity. 
%It will not however be the purpose of this work to complete the supersymmetry at other orders in the expansion.
%This is, as mentioned, a complicated task and also depends on each particular desired order in the expansion. 
At any desired order in the expansion in the Seeley-De Witt coefficients, one 
can construct  
generalized bosonic actions based on the Dirac operator of $\mathcal{N} = 1$ supergravity and the effective actions one is able to construct are the ones one should supersymmetrize, for 
each particular case, to get an effective supergravity spectral action.  Then,       
we present the ``bosonic action'' up to the $a_4$ term. 
Finally, Sec. 4 is devoted to conclusions and 
outlook. 

\section{Spectral action of pure gravity}

Instead of the well known geometry of space-time, the basic data of noncommutative geometry consists of an involutive 
algebra $\mathcal{A}$ of operators in a Hilbert space $\mathcal{H}$, which plays the role of the algebra of coordinates,  a self-adjoint operator of Dirac type $\mathcal{D}$ in $\mathcal{H}$ 
which plays the role of the 
inverse line element. A fundamental principle in the noncommutative approach is that the usual emphasis on points in space-time is replaced by the spectrum of the operator $\mathcal{D}$.
An operator is of Dirac type if its square is of Laplace type. Locally such operators can be expressed as
\begin{equation}
D = -(g^{\mu \nu}\nabla_\mu \nabla_\nu + E), \label{eq1}
\end{equation} 
\noindent for a unique endomorphism $E$ acting on vector bundles of $\mathcal{M}$. The spectral triple ($\mathcal{A}$,$\mathcal{H}$,$\mathcal{D}$) encodes the geometry 
of every noncommutative space. A Riemannian spin manifod $\mathcal{M}$ is completely characterized by the 
algebra of smooth functions on $\mathcal{M}$, $\mathcal{A} = C^\infty (\mathcal{M})$, the Hilbert space of square integrable spinors, $\mathcal{H} = L^2 (\mathcal{M},S)$ and the Dirac operator $\mathcal{D}$ 
of the Levi-Civita spin connection. On Riemannian manifolds $\mathcal{D}$ is an elliptic operator. The selfadjointness and ellipticity of $\mathcal{D}$ is essential for the construction of 
$(\mathcal{A},\mathcal{H},\mathcal{D})$. 
The spectral action principle states that the physical action depends only on the spectrum of the Dirac operator. These ideas were the origin of the spectral action given in  Ref. \citen{Connes}.
The bosonic part of the spectral action is
\begin{equation}
S = Tr{ \left [ f \left( \frac{D}{\Lambda} \right) \right]}. \label{eq2}
\end{equation}   
\noindent For a specific choice of the cutoff function $f$, the spectral action (\ref{eq2}) is expressed up to the first three terms of its asymptotic expansion \cite{Cham_Connes}
\begin{equation}
S = Tr{ \left[ f \left( \frac{D}{\Lambda} \right) \right]} \sim 2\Lambda^4 f_4 a_0 + 2\Lambda^2f_2 a_2 + f_0 a_4 ,\label{eq3}
\end{equation}   
\noindent where $f_4 = \int_0^\infty f(u) u^3 du$,$f_2 = \int_0^\infty f(u)udu$, $f_0 = f(0)$, and the $a_n$ are the Seeley-Dewitt coefficients of the heat kernel expansion of $D$. Every $a_n$ is function 
of geometric invariants of order $n$ constructed from $E$, the field strength $\Omega_{\mu \nu}$ and the Riemann tensor \cite{Vassi}. The relevant $a_n$'s are
{\setlength{\arraycolsep}{.2pt}
\begin{eqnarray}
&& a_0 = \frac{1}{4\pi^2}\int d^4x \sqrt{g} ,  \label{eq4} \\ \nonumber
&& a_2 = \frac{1}{16\pi^2} \int d^4x  \sqrt{g} Tr{(E + \frac{1}{6}R)} , \\ \nonumber
&& a_4 = \frac{1}{16\pi^2}\frac{1}{360} \int d^4x \sqrt{g} Tr{ (12 R_{; \mu}^{~\mu} + 5R^2 -2R_{\mu \nu}R^{\mu \nu}} \\ \nonumber
&& + 2R_{\mu \nu \rho \sigma}R^{\mu \nu \rho \sigma} + 60RE + 180E^2 + 60E_{; \mu}^{~\mu} + 30\Omega_{\mu \nu}\Omega^{\mu \nu}). 
\end{eqnarray}
For the gravitational Dirac operator $\mathcal{D}$ we have
{\setlength{\arraycolsep}{.2pt}
\begin{eqnarray}
& &\mathcal{D} = e^\mu_{~a}\gamma^a(\partial_\mu + \tilde{\omega}_\mu)~,\label{eq5} \\ \nonumber
& &E = -\frac{1}{4}R~,\\ \nonumber
& &\Omega_{\mu \nu} = \frac{1}{4}R_{\mu \nu}^{~~ab}\gamma_{ab}~,
\end{eqnarray}
\noindent where $	\tilde{\omega}_\mu$ is the spin connection on $\mathcal{M}$, 
$~\tilde{\omega}_\mu = \frac{1}{4}\tilde{\omega}_\mu^{~ab}\gamma_{ab}~$ with $~\tilde{\omega}_\mu~$  related to $~e_\mu^{~a}~$ 
by the vanishing of the covariant derivative $\nabla_\mu e_\nu^{~a} =0$,
this allows us to express the $~\tilde{\omega}_\mu^{~ab}~$ as functions of the tetrads as in standard 
Einstein tetradic gravity, applying the Riemannian torsion free condition. The coefficients (\ref{eq4}) take the form 

\begin{eqnarray}
& &a_0 = \frac{1}{4\pi^2}\int{d^4x \sqrt{g}}~, \\ \nonumber
& &a_2 = -\frac{1}{48\pi^2}\int d^4x\sqrt{g} R~, \\ \nonumber
& &a_4 = \frac{1}{4\pi^2}\frac{1}{360}\int d^4x\sqrt{g}~
(-18C_{\mu \nu \rho \sigma}C^{\mu \nu \rho \sigma} + 11R^*R^*)~,
\end{eqnarray}
\noindent where $C_{\mu \nu \rho \sigma}$ is the Weyl tensor of conformal gravity $
C_{\mu \nu \rho \sigma}C^{\mu \nu \rho \sigma} = R_{\mu \nu \rho \sigma}R^{\mu \nu \rho \sigma}
- 2R_{\mu \nu}R^{\mu \nu} + \frac{1}{3}R^2$ and the Euler characteristic $\chi_E$ is given by, 
$\chi_E = \frac{1}{32\pi}\int{d^4x \sqrt{g} R^*R^*}$ with $
R^*R^* = R_{\mu \nu \rho \sigma}R^{\mu \nu \rho \sigma} - 4R_{\mu \nu}R^{\mu \nu} + R^2$, the Gauss-Bonnet topological invariant.  

For this particular Dirac operator the spectral action is
\begin{equation}
S = \int d^4 x \sqrt{g} \left\{ \alpha + \beta R + \gamma (-18C_{\mu \nu \rho \sigma}C^{\mu \nu \rho \sigma} + 11R^*R^*) \right\}~,\label{grav}
\end{equation}
\noindent where $\alpha$, $\beta$ and $\gamma$ are constants. The spectral action gives the Hilbert-Einstein action with corrections. The action above is of 
particular interest, because this same expression has been considered as a good 
candidate for a renormalizable and ghost free theory of gravity \cite{Lu_Pope1,Lu_Pope2}. In the next section 
we consider the Dirac operator of $\mathcal{N}$ = 1 Euclidean supergravity and calculate the Seeley-Dewitt coefficients associated with its spectral action. The calculation of this action 
gives a ``generalized effective bosonic action'' being the natural supersymmetric extension of the spectral action for gravity, not including the appropriate matter fermionic terms that we would need to 
complete the supersymmetry. A procedure that one should perform at any desired order in the expansion due to the fact that we do not know a general expression for the ``matter fermionic'' 
supersymmetric 
action corresponding to the spectral action \cite{17}. As mentioned in the introduction, it has already been shown that under certain conditions \cite{Vancea1,Vancea2} the 
eigenvalues of the supergravity Dirac operator we use can be considered as observables of Euclidean supergravity. 

\section{Spectral action and Euclidean Supergravity}

Let $\mathcal{M}$ be a compact Riemannian spin manifold without boundary in four dimensions with metric $g_{\mu \nu} = e_\mu^{~a}e_{\nu a}$, the tetrad fields are labeled with greek space-time
 and latin internal indices respectively. The Dirac operator $\mathcal{D}$ given by the spin connection in (\ref{eq5}), is an elliptic operator on $\mathcal{M}$ and is formally selfadjoint on 
 $\mathcal{H}$. Because  $\mathcal{M}$ is compact, $\mathcal{D}$ admits a discrete spectrum of real eigenvalues and a complete set of eigenspinors $\mathcal{D} \psi^n = \lambda^n \psi^n$. 
 The $\lambda^n$'s define a discrete family of real valued functions on the phase space of smooth tetrad fields and as it was shown in Ref. \citen{Rovelli}, 
these eigenvalues are invariant under diffeomorphisms of $\mathcal{M}$ and under rotations of the tetrad fields, so they form a set of observables for general 
relativity. These ideas were extended in Ref. \citen{Vancea1,Vancea2} to achieve the geometric construction of Euclidean supergravity. This involving a supersymmetric partner of the graviton, 
the gravitino, and also imposing local supersymmetric invariance. The gravitino is represented by a Euclidean spinor vector $\psi_\mu^a$ defined by a Majorana condition, $\bar{\psi} = \psi^T C$.
The phase space will be the space of all pairs 
$(e,\psi)$ that are solution of the equations of motion modulo gauge transformations, which are the ones needed in the non supersymmetric case plus the transformations of local 
$\mathcal{N}$=1 supersymmetry. The supersymmetric Dirac operator $D_{SG}$ is given by 
\begin{equation}
D_{SG} = i\gamma^a e^\mu_a[\partial_\mu + 
(\tilde{\omega}_{\mu bc} + K_{\mu bc})\sigma^{bc}]~.\label{eq8}
\end{equation}
The difference between $D_{SG}$ and $\mathcal{D}$ is the additional $\psi$ dependent term \cite{Nieuwen,Deser},
$K_{\mu ab} = -\frac{1}{4}(\bar{\psi}_\mu \gamma_b \psi_a  - \bar{\psi}_a \gamma_\mu \psi_b  +  \bar{\psi}_b \gamma_a \psi_\mu )$. $D_{SG}$ is an elliptic operator defined on the full 
spin bundle and it is possible to define an inner product such that $D_{SG}$ is formally selfadjoint \cite{Vancea1}. Here we calculate the Seeley DeWitt coefficients of the square of $D_{SG}$. 
The square of this Dirac operator can be expressed in the form of (\ref{eq1}) with 
{\setlength{\arraycolsep}{.2pt}
\begin{eqnarray}
& & E =  -\frac{1}{4}R - \frac{1}{4}\nabla_\mu (\bar{\psi}^\mu \gamma_\nu \psi^\nu) + \frac{1}{16}\bar{\psi}_\alpha \gamma^\alpha \psi_\beta \bar{\psi}^\nu \gamma_\nu \psi^\beta  \label{eq9} \\ \nonumber 
& & + \frac{1}{32}\bar{\psi}^\nu \gamma^\alpha \psi_\beta \bar{\psi}_\alpha \gamma^\beta \psi_\nu - \frac{1}{64}\bar{\psi}_\nu \gamma_\alpha \psi_\beta \bar{\psi}^\nu \gamma^\alpha \psi^\beta ,
\end{eqnarray}
}
\noindent where $R$ is the curvature scalar of standard general relativity. The field strength is 
\begin{eqnarray}
&&\Omega_{\mu \nu} = \frac{1}{4}R_{\mu \nu}^{~~ab} \gamma_{ab} \label{eq10} \\  \nonumber
&&+ \frac{1}{16}[ \bar{\psi}_\mu \gamma^\sigma \psi^a  \bar{\psi}_\nu \gamma^b \psi_\sigma  - \bar{\psi}_\mu \gamma^\sigma \psi^a  \bar{\psi}_\sigma \gamma_\nu \psi^b 
+ \bar{\psi}_\mu \gamma^\sigma \psi^a  \bar{\psi}^b \gamma_\sigma \psi_\nu\\ \nonumber
&&-\bar{\psi}^a \gamma_\mu \psi^\sigma  \bar{\psi}_\nu \gamma^b \psi_\sigma +\bar{\psi}^a \gamma_\mu \psi^\sigma  \bar{\psi}_\sigma \gamma_\nu \psi^b 
 - \bar{\psi}^a \gamma_\mu \psi^\sigma  \bar{\psi}^b \gamma_\sigma \psi_\nu  \\ \nonumber
 &&+ \bar{\psi}^\sigma \gamma^a \psi_\mu  \bar{\psi}_\nu \gamma^b \psi_\sigma - \bar{\psi}^\sigma \gamma^a \psi_\mu  \bar{\psi}_\sigma \gamma_\nu \psi^b 
 + \bar{\psi}^\sigma \gamma^a \psi_\mu  \bar{\psi}^b \gamma_\sigma \psi_\nu ] \gamma_{ab} \\ \nonumber
 &&- \frac{1}{4}\nabla_\mu ( \bar{\psi}_\nu \gamma^b \psi^a  - \bar{\psi}^a \gamma_\nu \psi^b  + \bar{\psi}^b \gamma^a \psi_\nu ) \gamma_{ab}  -  (\mu \leftrightarrow \nu). \\ \nonumber
 \end{eqnarray}
By this means, and in a similar procedure as in pure gravity, we get the spectral action related to this particular Dirac operator and the constrained geometry defined by it. 
The first non constant term is
\begin{eqnarray}
&& a_2 =  -\frac{1}{48\pi^2}\int d^4xe ( R - \frac{1}{4}\bar{\psi}_\alpha \gamma^\alpha \psi_\beta \bar{\psi}^\nu \gamma_\nu \psi^\beta  \\  \nonumber
&& - \frac{1}{8} \bar{\psi}^\nu \gamma^\alpha \psi_\beta \bar{\psi}_\alpha \gamma^\beta \psi_\nu +  \frac{1}{16}\bar{\psi}_\nu \gamma_\alpha \psi_\beta \bar{\psi}^\nu \gamma^\alpha \psi^\beta  ).
\end{eqnarray}
The term $a_4$ is a combination of terms quadratic in $\psi$ and non trivial interactions between the graviton and the gravitino
{\setlength{\arraycolsep}{.2pt}
\begin{eqnarray}
&& a_4 =  \frac{1}{4\pi^2}\frac{1}{360}\int d^4x e [-18C_{\mu \nu \rho \sigma}C^{\mu \nu \rho \sigma} + 11R^*R^*  \label{eq13} \\ \nonumber
&& -14 R_{\mu \nu \rho \sigma} \Phi^{\mu \nu \rho \sigma} (\psi) - 106R_{\mu \nu} \Sigma^{\mu \nu} (\psi) +10R \Gamma (\psi) \\ \nonumber
&& -7 \Phi_{\mu \nu \rho \sigma}(\psi)\Phi^{\mu \nu \rho \sigma} (\psi) - 62\Sigma_{\mu \nu} (\psi)\Sigma^{\mu \nu}(\psi) + 5\Gamma^2(\psi)].
\end{eqnarray} 
}
The cuadratic terms appearing in (\ref{eq13}) are given by
{\setlength{\arraycolsep}{.2pt}
\begin{eqnarray}
&& \Phi_{\mu \nu \rho \sigma} (\psi)  =  \frac{1}{4}\nabla_\mu (\bar{\psi}_\rho \gamma_\sigma \psi_\nu 
+ \bar{\psi}_\rho \gamma_\nu \psi_\sigma + \bar{\psi}_\nu \gamma_\rho \psi_\sigma) \\ \nonumber
%& -\frac{1}{4}\nabla_\nu (\bar{\psi}_\rho \gamma_\sigma \psi_\mu + \bar{\psi}_\rho \gamma_\mu \psi_\sigma + \bar{\psi}_\mu \gamma_\rho \psi_\sigma) \\ \nonumber
&&+ \frac{1}{16}(\bar{\psi}_\rho \gamma_\alpha \psi_\mu \bar{\psi}^\alpha \gamma_\sigma \psi_\nu + \bar{\psi}_\rho \gamma_\alpha \psi_\mu \bar{\psi}^\alpha \gamma_\nu \psi_\sigma \\ \nonumber
&& ~~~~~+ \bar{\psi}_\rho \gamma_\alpha \psi_\mu \bar{\psi}_\nu \gamma^\alpha \psi_\sigma + \bar{\psi}_\rho \gamma_\mu \psi_\alpha \bar{\psi}^\alpha \gamma_\sigma \psi_\nu \\ \nonumber
&& ~~~~~+ \bar{\psi}_\rho \gamma_\mu \psi_\alpha \bar{\psi}^\alpha \gamma_\nu \psi_\sigma 
+ \bar{\psi}_\rho \gamma_\mu \psi_\alpha \bar{\psi}_\nu \gamma^\alpha \psi_\sigma \\ \nonumber
&& ~~~~~+ \bar{\psi}_\mu \gamma_\rho \psi_\alpha \bar{\psi}^\alpha \gamma_\sigma \psi_\nu + \bar{\psi}_\mu \gamma_\rho \psi_\alpha \bar{\psi}^\alpha \gamma_\nu \psi_\sigma \\ \nonumber
&& ~~~~~+ \bar{\psi}_\mu \gamma_\rho \psi_\alpha \bar{\psi}_\nu \gamma^\alpha \psi_\sigma) - (\mu \leftrightarrow \nu),
\end{eqnarray}
}
{\setlength{\arraycolsep}{.2pt}
\begin{eqnarray}
&& \Sigma_{\mu \nu} (\psi) =  \frac{1}{2}\nabla_\mu (\bar{\psi}_\nu \gamma^\alpha \psi_\alpha)\nonumber\\  
&& - \frac{1}{4}\nabla_\alpha (\bar{\psi}_\nu \gamma^\alpha \psi_\mu + \bar{\psi}_\nu \gamma_\mu \psi^\alpha + \bar{\psi}_\mu \gamma_\nu \psi^\alpha) \\ \nonumber
&& + \frac{1}{8}(\bar{\psi}_\nu \gamma_\beta \psi_\mu \bar{\psi}^\beta \gamma^\alpha \psi_\alpha + \bar{\psi}_\nu \gamma_\mu \psi_\beta \bar{\psi}^\beta \gamma^\alpha \psi_\alpha \\ \nonumber
&& ~~~+ \bar{\psi}_\mu \gamma_\nu \psi_\beta \bar{\psi}^\beta \gamma^\alpha \psi_\alpha)  \\ \nonumber
&& - \frac{1}{16}(\bar{\psi}_\nu \gamma_\beta \psi_\alpha \bar{\psi}^\beta \gamma^\alpha \psi_\mu + \bar{\psi}_\nu \gamma_\beta \psi_\alpha \bar{\psi}^\beta \gamma_\mu \psi^\alpha \\ \nonumber
&& ~~~~+ \bar{\psi}_\nu \gamma_\beta \psi_\alpha \bar{\psi}_\mu \gamma^\beta \psi^\alpha + \bar{\psi}_\nu \gamma_\alpha \psi_\beta \bar{\psi}^\beta \gamma^\alpha \psi_\mu \\ \nonumber
&& ~~~~+ \bar{\psi}_\nu \gamma_\alpha \psi_\beta \bar{\psi}^\beta \gamma_\mu \psi^\alpha + \bar{\psi}_\nu \gamma_\alpha \psi_\beta \bar{\psi}_\mu \gamma^\beta \psi^\alpha \\ \nonumber
&& ~~~~+ \bar{\psi}_\alpha \gamma_\nu \psi_\beta \bar{\psi}^\beta \gamma^\alpha \psi_\mu + \bar{\psi}_\alpha \gamma_\nu \psi_\beta \bar{\psi}^\beta \gamma_\mu \psi^\alpha \\ \nonumber
&& ~~~~+ \bar{\psi}_\alpha \gamma_\nu \psi_\beta \bar{\psi}_\mu \gamma^\beta \psi^\alpha),
\end{eqnarray}
}
\begin{eqnarray}
&& \Gamma (\psi) = \nabla_\mu (\bar{\psi}^\mu \gamma^\nu \psi_\nu) 
- \frac{1}{4}\bar{\psi}_\alpha \gamma^\alpha \psi_\beta \bar{\psi}^\nu \gamma_\nu \psi^\beta \label{eq16}\\ \nonumber
&& - \frac{1}{8}\bar{\psi}^\nu \gamma^\alpha \psi_\beta \bar{\psi}_\alpha \gamma^\beta \psi_\nu
+ \frac{1}{16}\bar{\psi}_\nu \gamma_\alpha \psi_\beta \bar{\psi}^\nu \gamma^\alpha \psi^\beta.
\end{eqnarray}
We recognize in the $a_2$ term the ``bosonic'' part of the $\mathcal{N}$=1 supergravity action written as a second-order formalism.  Now, as is well known, we can write it in terms of the 
curvature scalar that is function of the spin connection including the torsion term
\begin{eqnarray}
&& R(e,\psi) = R - \frac{1}{4}\bar{\psi}_\alpha \gamma^\alpha \psi_\beta \bar{\psi}^\nu \gamma_\nu \psi^\beta  
- \frac{1}{8} \bar{\psi}^\nu \gamma^\alpha \psi_\beta \bar{\psi}_\alpha \gamma^\beta \psi_\nu \\ \nonumber
&& ~~~~~~~~~~~~ + \frac{1}{16}\bar{\psi}_\nu \gamma_\alpha \psi_\beta \bar{\psi}^\nu \gamma^\alpha \psi^\beta  .
\end{eqnarray}        
Having modified the Dirac operator in a consistent way, summing the torsion term, the spectral action includes the geometric part of the $\mathcal{N}$= 1 supergravity action. 
The full supersymmetric action is not known. There is not a general ``matter fermionic'' supersymmetric partner of (\ref{eq2}), in particular not one constructed with the Dirac operator (\ref{eq8}). 
It is however possible, as we show, to provide at any desired order in the expansion of the action in the Seeley-DeWitt coefficients a ``generalized effective bosonic action'' and at the order 
of interest one can, in principle based in our result, supersymmetrize that effective particular bosonic action.     
The bosonic action (up to $a_4$) is then a higher order theory represented by the spectral action  
\begin{eqnarray}
&& S =  Tr [f(\frac{D_{SG}}{\Lambda})]  \label{18}\\ \nonumber
&& = \int d^4x e [ \alpha + \beta R(e,\psi)]  \\ \nonumber
&& + \gamma \int d^4x e[-18C_{\mu \nu \rho \sigma}C^{\mu \nu \rho \sigma} + 11R^*R^*  \\ \nonumber
&& -14 R_{\mu \nu \rho \sigma} \Phi^{\mu \nu \rho \sigma} (\psi) - 106R_{\mu \nu} \Sigma^{\mu \nu} (\psi) +10R \Gamma (\psi) \\ \nonumber
&& -7 \Phi_{\mu \nu \rho \sigma}(\psi)\Phi^{\mu \nu \rho \sigma} (\psi) - 62\Sigma_{\mu \nu} (\psi)\Sigma^{\mu \nu}(\psi) + 5\Gamma^2(\psi)],
\end{eqnarray}   
\noindent with $\alpha$, $\beta$, and $\gamma$ constants. 
It is of interest to notice that the spectral action (\ref{eq2}) for $D = \mathcal{D}^2$ is a theory of gravitation. It includes in a natural way  
the Einstein-Hilbert action.  A general higher order theory of gravity suffers from ghosts, a scalar and a spin-2 mode, both massive. However, the particular theory of gravitation that emerges 
in the non commutative geometry framework, up to the $a_4$ term, is of the type  
\begin{equation}
I = \frac{1}{2\kappa^2} \int d^4x e \left( R - 2\Lambda 
+ \frac{1}{2}\alpha C_{\mu \nu \rho \alpha}C^{\mu \nu \rho \alpha} \right) , \label{eq18}
\end{equation}
\noindent  this form of the theory, without a cosmological constant, was considered first in Ref. \citen{Stelle} and it was argued that it is renormalizable. 
It was then reconsidered in Ref. \citen{Lu_Pope1} including $\Lambda$ 
because in this case the scalar mode is absent and for a special value of $\alpha$ in terms of $\Lambda$, the massive spin-2 mode also disappears, leaving a theory consisting only of a massless 
graviton and is possibly renormalizable. The spectral action we propose (\ref{18}) is the ``bosonic'' part of the supergravity spectral action, that one should construct in a 
supersymmetric procedure at each order in the expansion in the Seeley-De Witt coefficients. 
It will correspond to the gravity action (\ref{eq18}) 
in Ref. \citen{Lu_Pope2}. The action (\ref{18}) gives, up to $a_2$, the usual simple supergravity by adding the Rarita-Schwinger action $\langle \Psi ,D_{SG} \Psi \rangle$. 
The supersymmetric action up to $a_4$ could be written 
in the form of an Einstein-Weyl supergravity without the scalar and vector auxiliary fields and it could also be studied in relation to its renormalizability. On the other hand,  
the Seeley-Dewitt coefficients have been calculated, in particular, when totally antisymmetric torsion is present and by these means the associated spectral action \cite{11,12,13,14,15,16}.
The motivations of doing so are based on some physical arguments, for instance the coincidence of geodesics in both manifolds, one in which there is torsion and another where it is absent.
 Torsion in supergravity is of a more general kind and there is no reason, physical or mathematical, for not considering its associated Dirac operator as we have done.  

\section{Discussions}

The spectral action allows us to construct a modified theory of gravity as well as the ``bosonic'' part of a generalized supergravity at the desired order in the Seeley-DeWitt coefficients. 
The fact that makes this possible is that simple supergravity and pure gravity can both be viewed as theories with well defined, mathematically consistent, Dirac operators.   
This is not, in general,  the case for extended supergravities. We were partially motivated by the result that the eigenvalues of the supergravity Dirac operator can, with certain constraints, be 
considered as observables \cite{Vancea1,Vancea2}. We were able to calculate the expansion up to $a_4$ of the spectral action (\ref{eq2}) for the supersymmetric Dirac operator (\ref{eq8}), and 
we have constructed a generalized bosonic effective action that provides at each order in the expansion of the Seeley-DeWitt coefficients an effective action that is the appropriate to be, 
in principle, supersymmetrized to get  an effective supergravity action. 
The gravity action (\ref{grav},\ref{eq18}) has been constructed with the pure gravity spin connection (\ref{eq5}). This gravity action was already known 
\cite{Lu_Pope2,Stelle} and may be considered to possibly be renormalizable. The corresponding supergravity action would be given by the supersymmetric completion of our action 
(\ref{18}). It is a matter of further work to search for the calculation of this effective supergravity theory and look for its possible renormalizability.

\section*{Acknowledgments}
O. Obreg\'on and M. Sabido are partially supported by CONACYT grants 62253, 135023 and DAIP 125/11. J. L. L\'opez was supported by CONACYT grant  43683.
This work is part of red PROMEP UABC-UAM-UGTO.

\end{document}